
\magnification = 1200
\overfullrule=0pt

\font\titlerm = cmr10 scaled\magstep 4
\font\titlerms = cmr7 scaled\magstep 4
\font\titlermss = cmr5 scaled\magstep 4
\font\titlei = cmmi10 scaled\magstep 4
\font\titleis = cmmi7 scaled\magstep 4 
\font\titleiss = cmmi5 scaled\magstep 4
\font\titlesy = cmsy10 scaled\magstep 4 
\font\titlesys = cmsy7 scaled\magstep 4
\font\titlesyss = cmsy5 scaled\magstep 4 
\font\titleit = cmti10 scaled\magstep 4

\def\titlefont{\def\rm{\fam0\titlerm}
\def\it{\fam\itfam\titleit}
\textfont0 = \titlerm
\scriptfont0 = \titlerms
\scriptscriptfont0 = \titlermss
\textfont1 = \titlei
\scriptfont1 = \titleis
\scriptscriptfont1 = \titleiss
\textfont2 = \titlesy
\scriptfont2 = \titlesys
\scriptscriptfont2 = \titlesyss
\textfont\itfam = \titleit
\rm}

\def\sectionfont{\def\rm{\fam0\tenrm}
\def\it{\fam\itfam\tenit}
\def\bf{\fam\bffam\tenbf}
\textfont0 = \tenrm
\scriptfont0 = \sevenrm
\scriptscriptfont0 = \fiverm
\textfont1 = \teni
\scriptfont1 = \seveni  \scriptscriptfont1=\fivei
\textfont2 = \tensy
\scriptfont2 = \sevensy
\scriptscriptfont2 = \fivesy
\textfont\itfam = \tenit  
\textfont\bffam = \tenbf
\rm}

\font\teenyfont = cmr5

\global\baselineskip = 1.2\baselineskip
\global\parskip = 4pt plus 0.3pt
\global\nulldelimiterspace = 0pt

\predisplaypenalty 1000


\def\endignore{}
\def\ignore #1\endignore{}

\newcount\dflag
\dflag = 0


\def\monthname{\ifcase\month
\or Jan \or Feb \or Mar \or Apr \or May \or June%
\or July \or Aug \or Sept \or Oct \or Nov \or Dec
\fi}




\def\endid{}
\def\id#1\endid{\number\day\ \monthname \number\year
\hfill #1}

\def\endtitle{}
\def\title#1\endtitle{\vskip.15in\titlefont
\global\baselineskip = 2\baselineskip
#1\vskip.3in
\baselineskip = 0.5\baselineskip\sectionfont}

\def\lblfoot{This work was supported by the Director, Office of Energy 
Research, Office of High Energy and Nuclear Physics, Division of High 
Energy Physics of the U.S. Department of Energy under Contract 
DE-AC03-76SF00098.}

\def\endauthors{}
\def\authors#1\endauthors{
#1\if\dflag = 0
\footnote{}{\noindent\lblfoot}\fi}

\def\endabstract{}
\def\abstract#1\endabstract{\vskip .2in%
\centerline{\sectionfont\bf Abstract}%
\vskip .1in%
\noindent#1%
\ifnum\dflag = 0
\footline = {\hfil}\pageno = 0
\vfill\eject
\pageno = 1\footline{\centerline{\sectionfont\folio}}
\fi\ifnum\dflag = 2
\footline = {\hfil}\pageno = 0
\vfill\eject
\fi}


\newcount\nsection
\newcount\nsubsection

\def\section#1{\global\advance\nsection by 1
\global\nsubsection = 0
\bigskip\noindent
\centerline{\sectionfont\bf\number\nsection.\ #1}
\nobreak\medskip\sectionfont\nobreak}

\def\subsection#1{\global\advance\nsubsection by 1
\bigskip\noindent
\centerline{\sectionfont \it \number\nsection.\number\nsubsection.\ #1}
\nobreak\vskip-0.1in\rm\nobreak}

\def\appendix#1#2{\bigskip\noindent%
\sectionfont \bf Appendix #1.\ #2
\nobreak\medskip\rm\nobreak}


\newcount\nref
\global\nref = 1

\def\ref#1#2{\xdef #1{[\number\nref]}
#1
\ifnum\nref = 1\global\xdef\therefs{\noindent[\number\nref] #2\ }
\else
\global\xdef\oldrefs{\therefs}
\global\xdef\therefs{\oldrefs\vskip.1in\noindent[\number\nref] #2\ }%
\fi%
\global\advance\nref by 1
}

\def\listrefs{\vfill\eject\section{References}\therefs}


\newcount\nfig
\global\nfig = 1

\def\fg#1\efig{\vskip .5in\noindent Fig.\ \number\nfig:\ #1%
\global\advance\nfig by 1}


\newcount\cflag
\newcount\nequation
\global\nequation = 1
\def\eqlabel{(1)}

\def\nexteqno{\ifnum\cflag = 0
\global\advance\nequation by 1
\fi
\global\cflag = 0
\xdef\eqlabel{(\number\nequation)}}

\def\lasteqno{\global\advance\nequation by -1
\xdef\eqlabel{(\number\nequation)}}

\def\label#1{\xdef #1{(\number\nequation)}
\ifnum\dflag = 1
{\escapechar = -1
\xdef\draftname{\teenyfont\string#1}}
\fi}

\def\clabel#1#2{\xdef\eqlabel{(\number\nequation #2)}
\global\cflag = 1
\xdef #1{\eqlabel}
\ifnum\dflag = 1
{\escapechar = -1
\xdef\draftname{\string#1}}
\fi}

\def\cclabel#1#2{\xdef\eqlabel{#2)}
\global\cflag = 1
\xdef #1{\eqlabel}
\ifnum\dflag = 1
{\escapechar = -1
\xdef\draftname{\string#1}}
\fi}


\def\eeq{}

\def\eqnn #1\eeq{$$ #1 $$}

\def\eq #1\eeq{\xdef\draftname{\ }
$$ #1
\eqno{\eqlabel \rlap{\ \draftname}} $$
\nexteqno}



\def\eol{& \eqlabel \rlap{\ \draftname} \crcr
\nexteqno
\xdef\draftname{\ }}

\def\eeol{& \eqlabel \rlap{\ \draftname}
\nexteqno
\xdef\draftname{\ }}

\def\eolnn{\cr
\global\cflag = 0
\xdef\draftname{\ }}


\def\eqa #1\eeq{\xdef\draftname{\ }
$$ \eqalignno{ #1 } $$
\global\cflag = 0}


\def\eg{{\it e.g.\/}}

\def\myinstitution{
    \centerline{\it Theoretical Physics Group}
    \centerline{\it Lawrence Berkeley Laboratory}
    \centerline{\it 1 Cyclotron Road}
    \centerline{\it Berkeley, California 94720}
}


\def\jref#1#2#3#4{{\it #1} {\bf #2}, #3 (#4)}

\def\NPB#1#2#3{\jref{Nucl.\ Phys.}{B#1}{#2}{#3}}
\def\PA#1#2#3{\jref{Physica}{#1A}{#2}{#3}}
\def\PLB#1#2#3{\jref{Phys.\ Lett.}{#1B}{#2}{#3}}

\def\PRD#1#2#3{\jref{Phys.\ Rev.}{D#1}{#2}{#3}}

\def\PRL#1#2#3{\jref{Phys.\ Rev.\ Lett.}{#1}{#2}{#3}}
\def\PRV#1#2#3{\jref{Phys.\ Rev.}{#1}{#2}{#3}}


\def\to{\mathop{\rightarrow}}


\def\frac#1#2{{{#1} \over {#2}}\,}  
\def\sfrac#1#2{{\textstyle\frac{#1}{#2}}}  


\def\Dsl{\hbox{/\kern-.6000em\rm D}} 



\def\scr#1{{\cal #1}}

\def\mybar#1{\kern 0.8pt\overline{\kern -0.8pt#1\kern -0.8pt}\kern 0.8pt}
\def\sla#1{\raise.15ex\hbox{$/$}\kern-.57em #1}
\def\Sla#1{\kern.15em\raise.15ex\hbox{$/$}\kern-.72em #1}

\def\roughly#1{\mathrel{\raise.3ex\hbox{$#1$\kern-.75em%
    \lower1ex\hbox{$\sim$}}}}


\def\tr{\mathop{\rm tr}}


\def\bra#1{\langle #1 |}
\def\ket#1{| #1 \rangle}



\hyphenation{ba-ry-on ba-ry-ons ano-ma-ly ano-ma-lies}

\def\al{\alpha}
\def\del{\delta}
\def\Del{\Delta}
\def\gam{\gamma}

\def\ep{\epsilon}

\def\Lam{\Lambda}

\def\Sig{\Sigma}

\def\ChPT{\raise.45ex\hbox{$\chi$}PT}

\def\hc{{\rm h.c.}}


\def\MeV{{\rm \ MeV}}
\def\GeV{{\rm \ GeV}}

\overfullrule=0pt

\def\chisim{$SU(3)_L \times SU(3)_R$}
\def\lchi{\Lam_\chi}
\def\nc{N_{\rm c}}


\id
LBL-34040
\endid
\rightline{CfPA-TH-93-10}

\title
\centerline{Decuplet Contributions to}
\centerline{Hyperon Axial Vector Form Factors}
\endtitle

\authors
\centerline{Markus A. Luty}
\vskip .1in
\myinstitution
\vskip .15in
\centerline{Martin White}
\vskip .1in
\centerline{{\it Center for Particle Astrophysics}}
\centerline{{\it 301 Le Conte Hall}}
\centerline{{\it University of California}}
\centerline{{\it Berkeley, CA 94720}}
\endauthors

\abstract
We consider the predictions of chiral perturbation theory for $SU(3)$
breaking in the axial vector form factor $g_1$ measured in semileptonic
hyperon decays.
We confirm that if only octet baryon intermediate states are included,
the non-analytic corrections are $\sim 100\%$.
These corrections are dominated by an $SU(3)$-symmetric wavefunction
renormalization, which explains the fact that the ``corrected''
predictions still fit the data well.
We argue that the large corrections are nonetheless strong evidence that
the chiral expansion is breaking down.
Following a recent suggestion of Jenkins and Manohar, we then include
contributions from decuplet baryon intermediate states.
Unlike these authors, we do not neglect the octet--decuplet mass
difference $\Delta$.
We find that the effects of $\Delta \ne 0$ significantly change the
pattern of corrections:
we still find that the decuplet corrections can cancel the large octet
contributions in a non-trivial way, but the corrections no longer favor
the $SU(6)$ values of the axial couplings.
We also argue that $D$ and $F$ axial couplings cannot be reliably
extracted from calculations which keep only the non-analytic corrections.
\endabstract


\section{Introduction}

Chiral perturbation theory (\ChPT) provides a rigorous framework for
extracting physical predictions from QCD as power series in the ``light''
quark current masses $m_u$, $m_d$, and $m_s$.
At lowest order, \ChPT\ predicts a large number of quantities in terms of
a few effective coupling constants.
Many of the resulting predictions are very successful.

Clearly, it is important to understand the size of the corrections
to the lowest order results, especially for quantities sensitive to $m_s$,
since experience with the chiral expansion suggests that the fractional
corrections to these quantities are of order
\eq
\frac{m_s}{\lchi} \sim 0.2,
\eeq
where $\lchi \sim 1 \GeV$ is the \ChPT\ expansion parameter.
Unfortunately, higher orders in the chiral expansion involve many
unknown effective couplings.\footnote{$^{\rm *}$}
{These effective coupling constants are related to QCD matrix elements which
are in principle computable (on the lattice, for example).
However, our present knowledge of the relevant matrix elements is rather
limited, and we conservatively regard the effective couplings as
undetermined parameters.}
However, there are nonanalytic corrections of order $m_s \ln m_s$
(and sometimes $m_s^{1/2}$ for processes involving baryons) which 
are computable in terms of the lowest-order couplings.
For sufficiently small values of $m_s$, these are the largest corrections.
While the nonanalytic corrections are not expected to be significantly
larger than the $O(m_s)$ contributions in the real world, the nonanalytic
corrections can be used to give an estimate of the expected size of
corrections.
In particular, if the nonanalytic corrections are large, then chiral
perturbation theory breaks down unless the $O(m_s)$ corrections cancel
the nonanalytic contributions.
Such a cancellation is unnatural, since it can occur only for special
values of the quark masses.
Thus, the calculation of the non-analytic contributions gives non-trivial
information about the chiral expansion, even if they cannot be used
to quantitatively predict the size of the corrections.

In ref.\ \ref\JMax{E. Jenkins and A. V. Manohar, \PLB{255}{558}{1991}.},
baryon chiral perturbation theory was reformulated in terms of an effective
lagrangian which includes the octet baryons as heavy fields
\ref\heavy{H. Georgi, \PLB{240}{447}{1990};
T. Mannel, W. Roberts, and Z. Ryzak, \NPB{368}{204}{1992}.},
and it was found that the leading nonanalytic corrections to the baryon
axial form factors were $\sim 100\%$.\footnote{$^\dagger$}
{An earlier calculation
\ref\Wise{J. Bijnens, H. Sonoda, and M. Wise, \NPB{261}{185}{1985}.}\ 
which found smaller corrections is incorrect.}
This is surprising in light of the fact that the lowest-order predictions
work to better than $20\%$.
In this paper, we point out that these corrections are dominated by an
$SU(3)$-singlet wavefunction renormalization, which explains why the
fit to the data including the corrections still works well.
However, we argue that the presence of these large corrections is
nonetheless strong evidence that the chiral expansion defined by this
effective lagrangian is breaking down.

In ref.\ \ref\JMdec{E. Jenkins and A. V. Manohar, \PLB{259}{353}{1991}.},
it was found that the large chiral symmetry breaking corrections discussed
above are largely cancelled by contributions from decuplet intermediate states.
From the point of view of the effective lagrangian, this looks like
an unnatural cancellation between two unrelated sectors of the theory.
However from the point of view of QCD the decuplet and octet states are
certainly related.
In fact, in the large-$\nc$ limit of QCD, the baryon spectrum
consists of $\sim \nc$ states with mass differences
$\sim \Lam_{\rm QCD} / \nc$, and the lowest-lying states have the quantum
numbers of the octet and decuplet baryons (for $\nc$ odd).
There are non-trivial relations between the different baryon multiplets
in this limit which reproduce the $SU(6)$ spin--flavor symmetry
relations between octet and decuplet baryon couplings
\ref\six{J.-L. Gervais and B. Sakita, \PRL{52}{87}{1984},
\PRD{30}{1795}{1984};
K. Bardakci, \NPB{243}{197}{1984};
R. Dashen and A. V. Mahohar, UCSD/PTH 93-16,18.}.
While these developments are very interesting, we will take a more
phenomenological point of view in this paper, including the decuplet
without assuming $SU(6)$ symmetry.

Including the decuplet in the chiral lagrangian is only justified if
the decuplet--octet mass splitting $\Del$ is small compared to the
\ChPT\ expansion parameter $\lchi$ (numerically, $\Del \simeq 300 \MeV$,
$\lchi \sim 1 \GeV$.)
In this limit, we can treat the decuplet as a nonrelativistic heavy field,
substantially simplifying the computation.
Our calculation differs from that of ref.\ \JMdec\ mainly in that we do not
make the approximation $\Del \ll m_K$.
We find that the effect of nonzero $\Del$ completely changes the pattern
of the non-analytic corrections:
the decuplet corrections can still cancel the large octet corrections
for some values of the couplings, but the corrections strongly
disfavor the $SU(6)$ values of the decuplet axial couplings.
We also argue that the axial coupling constants cannot be accurately
determined if only the non-analytic corrections are included.

This paper is organized as follows.
In section 2, we briefly review the effective lagrangian formalism we
will use.
In section 3, we present the results of our calculation including only
octet intermediate states.
In section 4, we present the results of including the decuplet.
Section 5 contains our conclusions.
Some information on the semileptonic decays used in this paper
is given in an appendix.

\section{Effective Lagrangian}

It has been known for some time that the low-energy theorems of
spontaneous chiral symmetry breaking can be encoded in an effective
lagrangian
\ref\effL{J. Schwinger, \PLB{24}{473}{1967};
S. Coleman, J. Wess, and B. Zumino, \PRV{117}{2239}{1969};
C. G. Callan, S. Coleman, J. Wess, and B. Zumino, \PRV{117}{2247}{1969};
S. Weinberg, \PA{96}{327}{1979}.}.
The lagrangian gives a systematic framework for investigating deviations
from the symmetry limit $m_u, m_d, m_s \to 0$.
In this section, we briefly review the effective lagrangian we use
to carry out the computation.
The notation and conventions we use are the same as those of
ref.\ \ref\ourvec{J. Anderson and M. A. Luty, LBL preprint LBL-33435, to
appear in {\it Phys.\ Rev.} {\bf D}.}.
The reader familiar with this formalism is urged to skip to section 3.

The effective lagrangian we will use includes the pseudoscalar meson
octet
\eq
\Pi = \frac 1{\sqrt 2}
\pmatrix{\frac 1{\sqrt 2}\pi^0 + \frac 1{\sqrt 6}\eta &
\pi^+ & K^+ \cr
\pi^- & -\frac 1{\sqrt 2} \pi^0 + \frac 1{\sqrt 6}\eta &
K^0 \cr
K^- & {\mybar K}^0 & -\frac 2{\sqrt 6} \eta \cr},
\eeq
the baryon octet
\eq
B = \pmatrix{
\frac 1{\sqrt 2} \Sigma^0 + \frac 1{\sqrt 6}\Lam &
\Sigma^+ & p \cr
\Sigma^- &
-\frac 1{\sqrt 2}\Sigma^0 + \frac 1{\sqrt 2}\Lam & n \cr
\Xi^- & \Xi^0 & -\frac 2{\sqrt 6}\Lam \cr},
\eeq
and the spin-$\frac 32$ decuplet.
The mesons are taken to transform under \chisim\ as
\eq
\xi \equiv e^{i\Pi / f}
\mapsto L \xi U^\dagger = U \xi R^\dagger,
\eeq
where the last equation defines $U$ as a function of $L$, $R$, and $\xi$.
The baryons are treated as heavy fields with four-velocity $v$, and
transform under \chisim\ as
\eq
B \mapsto U B U^\dagger.
\eeq
The lowest-order terms in the effective lagrangian involving the octet
baryons are
\eq
\scr L_{\rm oct} = \tr(\mybar B iv\cdot\partial B)
+ 2D \tr\left(\mybar B s^\mu \{ A_\mu, B \}\right)
+ 2F \tr\left(\mybar B s^\mu [ A_\mu, B ]\right) + \cdots.
\eeq

The decuplet fields are represented by a Rarita--Schwinger field $T$
with both vector and spinor indices.
We work in the limit where the octet--decuplet mass splitting $\Del$
is small compared to $\lchi$, so we can treat $T$ as a heavy field with the
same velocity $v$ as the baryon octet.
(We neglect splitting within the octet and decuplet induced by quark
masses, since these are higher order in the chiral expansion.)
The physical spin-$\frac 32$ components are projected out by imposing
the constraints \JMdec
\eq
v^\mu T_\mu = s^\mu T_\mu = 0,
\eeq
where $s$ is the spin matrix \JMax.

The decuplet fields can be represented as a completely symmetric
3-index tensor transforming under \chisim\ as
\eq
T^{jkl} \mapsto {U^j}_m {U^k}_n {U^l}_p T^{mnp}.
\eeq
The relevant terms in the lagrangian involving decuplet fields are
\eq
\label\decl
\scr L_{\rm dec} = - \mybar T^\mu iv\cdot\partial T_\mu
+ \Del\, \mybar T^\mu T_\mu
+ 2\scr H\, \mybar T^\mu s\cdot A T_\mu
+ \scr C\, (\mybar T^\mu A_\mu B + \hc),
\eeq
where $A$ is the axial current formed from pion fields.
We have used an abbreviated notation in which $SU(3)$ indices are
suppressed:
\eq
\mybar T A T \equiv \mybar T_{jkl} {A^j}_m T^{mkl}, \qquad
\mybar T A B \equiv \mybar T_{jkl} {A^j}_m {B^k}_n \ep^{lmn}.
\eeq

\section{Octet Corrections}

In this section, we will discuss the corrections which involve only
octet intermediate states.
Our calculation differs from that of ref.\ \JMax\ only in that we keep
$m_\pi \ne 0$.
The $\pi$ corrections are expected to be $\sim 20\%$ of the $K$ and $\eta$
corrections, but omitting the $\pi$ contributions systematically
{\it increases} the predicted $SU(3)$ breaking, so we include them here.

We write
\eq
g_1^{abc}(0) = \al_{ab}^c + \frac 1{16\pi^2 f^2} \beta_{ab}^c,
\eeq
where the lowest-order results are
\eq
\label\loword
\eqalign{
\al_{pn}^{1 + i2} &= D + F, \cr
\al_{\Lam\Sig^-}^{1 + i2} &= \frac 2{\sqrt 6} D, \cr
\al_{p\Lam}^{4 + i5} &= -\frac 1{\sqrt 6} (D + 3F), \cr
\al_{\Lam\Xi^-}^{4 + i5} &= -\frac 1{\sqrt 6} (D - 3F), \cr
\al_{n\Sig^-}^{4 + i5} &= D - F, \cr
\al_{\Sig^0\Xi^-}^{4 + i5} &= \sqrt 2\al_{\Sig^+\Xi^0}^{4 + i5}
= \frac 1{\sqrt 2} (D + F). \cr}
\eeq
The leading chiral corrections are
\def\fpi{\,m_\pi^2 \ln\frac{m_\pi^2}{\mu^2}}
\def\fk{\,m_K^2 \ln\frac{m_K^2}{\mu^2}}
\def\feta{\,m_\eta^2 \ln\frac{m_\eta^2}{\mu^2}}
\eqa
\label\firstcorr
\beta_{pn}^{1 + i2} &=
-(D + F)(2 D^2 + 4 D F + 2 F^2 + 1) \fpi\eolnn
&\qquad - \frac 16 (13 D^3 - D^2 F + 3 D + 3 D F^2 + 3 F + 33 F^3)  
\fk\eolnn
&\qquad - \frac 13 (D + F) (D - 3F)^2 \feta, \eol
\beta_{\Lam\Sig^-}^{1 + i2} &=
-\frac 2{3\sqrt 6} D (7 D^2 + 3 F^2 + 3) \fpi\eolnn
&\qquad - \frac 1{\sqrt 6} D (3 D^2 + 13 F^2 + 1) \fk\eolnn
&\qquad - \frac 4{3\sqrt 6} D^3 \feta, \eeol
\eeq
%
%
\eqa
\beta_{p\Lam}^{4 + i5} &=
\frac 3{8\sqrt 6}
(3 D^3 + 27 D^2 F + D + 25 D F^2 + 3 F + 9 F^3) \fpi\eolnn
&\qquad +\frac 1{12 \sqrt 6}
(31 D^3 + 15 D^2 F + 9 D + 9 D F^2 + 27 F + 297 F^3) \fk\eolnn
&\qquad + \frac 1{24 \sqrt 6}
(D + 3 F)(19 D^2 - 30 D F + 27 F^2 + 9) \feta, \eol
\beta_{\Lam\Xi^-}^{4 + i5} &=
\frac 3{8 \sqrt 6}(3 D^3 - 27 D^2 F + D + 25 D F^2 - 3 F - 9 F^3)  
\fpi\eolnn
&\qquad + \frac 1{12\sqrt 6}
(31 D^3 - 15 D^2 F + 9 D + 9 D F^2 - 27 F - 297 F^3) \fk\eolnn
&\qquad + \frac 1{24\sqrt 6}
(D - 3F)(19 D^2 + 30 D F + 27 F^2 + 9)\feta, \eol
\beta_{n\Sig^-}^{4 + i5} &=
-\frac 1{24}(35 D^3 + 23 D^2 F + 9 D + 33 D F^2 - 9 F - 123  
F^3)\fpi\eolnn
&\qquad - \frac 1{12}
(31 D^3 - 53 D^2 F + 9 D + 57 D F^2 - 9 F - 51 F^3)\fk\eolnn
&\qquad - \frac 1{24}
(D - F)(11 D^2 - 6 D F + 27 F^2 + 9) \feta, \eol
\label\lastcorr
\beta_{\Sig^0\Xi^-}^{4 + i5} &=
-\frac 1{24\sqrt 2}
(35 D^3 - 23 D^2 F + 9 D + 33 D F^2 + 9 F + 123 F^3) \fpi\eolnn
&\qquad -\frac 1{12\sqrt 2}
(31 D^3 + 53 D^2 F + 9 D + 57 D F^2 + 9 F + 51 F^3) \fk\eolnn
&\qquad - \frac 1{24\sqrt 2}
(D + F)(11 D^2 + 6 D F + 27 F^2 + 9)\feta.\eeol
\eeq
Here $\mu$ is an arbitrary renormalization scale.
The $\mu$ dependence of these results is cancelled by the $\mu$  
dependence of $O(m_s)$ terms in the effective lagrangian such as
\eq
\label\higher
\frac {c(\mu)}{\lchi}\,
\tr\left[ \mybar B (\xi^\dagger m_q \xi^\dagger + \hc) s \cdot A B \right],
\eeq
where $m_q$ is the quark mass matrix.
If we take $\mu \simeq \lchi$, there are no large logarithms in the
higher order coefficients, and near the chiral limit the correction
is dominated by the logarithmically enhanced terms computed above.
In the real world these logarithms are not very large, but we expect
that the logarithmic terms will give some indication of the actual  
size of the corrections, as discussed in the introduction.

The corrections to individual form factors are all larger than $80\%$,
in agreement with the results of ref.\ \JMax.
Closer inspection of the results of this calculation reveals that
the largest part of the corrections comes from the $SU(3)$-invariant
part of the wavefunction renormalization.
This contribution can be written
\eq
\del g_1^{abc}(0) = \al_{ab}^c \del Z,
\eeq
where
\eq
\del Z = \frac 18 (5 D^2 + 9 F^2) \left[3 \fpi + 4 \fk + \feta\right].
\eeq
The logarithmically-enhanced wavefunction renormalization must be
positive on general grounds
\ref\Collins{See \eg\ J. C. Collins, {\it Renormalization}
(Cambridge University Press, 1984), p. 192.},
and therefore we know that there must be positive $SU(3)$-invariant
piece.
The surprise is that this is by far the most important correction.
(If we remove this contribution, the largest correction is $\sim 50\%$,
with all other corrections less than $\sim 25\%$.)

This contribution can be formally removed from the chiral expansion by
defining the baryon fields
\eq
B' \equiv (1 + \del Z)^{-1/2} B.
\eeq
We can then write the terms in the lagrangian involving two baryon fields
as
\eqa
\scr L_B &= \tr(\mybar B iv\cdot\nabla B)
+ \sum_j c_j \scr O_j(B) \eolnn
&= (1 + \del Z) \tr(\mybar B iv\cdot\nabla B)
+ \sum_j c'_j \scr O_j(B'), \eeol
\eeq
where $c'_j = (1 + \del Z) c_j$.
If we now expand in terms of the coefficients $c'_j$ treating $\del Z$ as
order $m_s \ln m_s$, the large wavefunction renormalization is
absorbed into a redefinition of the chiral couplings.

Since wavefunction renormalization is universal for all amplitudes
with two external baryon lines, one might think that the resummation
discussed above shows that the large wavefunction correction is
``trivial,'' simply rescaling the couplings of the effective lagrangian
and leaving relations among observables intact.
However, $\del Z$ depends on the quark masses, and chiral symmetry
relates this dependence to physical quantities.
In particular, the $\mu$ dependence of $\del Z$ is compensated by the
$\mu$ dependence of the term
\eq
\frac{c(\mu)}{\lchi} \tr(\xi^\dagger m_q \xi^\dagger + \hc)
\tr(\mybar B iv\cdot\nabla B).
\eeq
If we chose $\mu$ to make $\del Z$ small, $c(\mu)$ will be large.
This leads to a breakdown of chiral perturbation theory, since this term
contributes to \eg\ $s$-wave pion--nucleon scattering.
While a full calculation would be required to demonstrate the breakdown
of chiral pertrubation theory, it is clear that there is no reason to
think that the large wavefunction renormalization corrections found
in this calculation are trivial.

\section{Decuplet Corrections}

In this section, we include the decuplet contributions to $g_1$.
We write
\eq
g_1^{abc}(0) = \al_{ab}^c + \frac 1{16\pi^2 f^2}
\Bigl[ \beta_{ab}^c + \scr C^2 \gamma_{ab}^c \Bigr],
\eeq
where $\gam$ contains the decuplet contributions, which are proportional
to the coupling $\scr C$ defined in eq.\ \decl.
We have
\def\gpi{G_1(m_\pi)}
\def\gk{G_1(m_K)}
\def\geta{G_1(m_\eta)}
\def\ggpi{G_2(m_\pi)}
\def\ggk{G_2(m_K)}
\def\ggeta{G_2(m_\eta)}
\eqa
\gam_{pn}^{1 + i2} &= -\frac 12(D + F) [4 \gpi + \gk]
- \frac{10}{81} \scr H [5\gpi + \gk] \eolnn
&\qquad\quad + \frac 29 [8(D + F)\ggpi +
(3D + F)\ggk], \eol
\gam_{\Lam\Sig^-}^{1 + i2} &= -\frac 1{6\sqrt 6} D
[11\gpi + 16\gk + 3\geta] \eolnn
&\qquad\quad - \frac 5{27\sqrt 6} \scr H [2\gpi + \gk] \eolnn
&\qquad\quad + \frac 2{9\sqrt 6} [(D + 6F)\ggpi
+ (8D + 12F)\ggk + 3D \ggeta], \eol
\gam_{p\Lam}^{4 + i5} &= \frac 1{4\sqrt 6}(D + 3F)[7\gpi + 3\gk]
+ \frac 5{9\sqrt 6} \scr H [2\gpi + \gk] \eolnn
&\qquad\quad + \frac 1{3\sqrt 6} [(11D + 3F)\ggpi
+ 3(D + F)\ggk], \eol
\gam_{\Lam\Xi^-}^{4 + i5} &= \frac 1{4\sqrt 6}(D - 3F)
[4\gpi + 5\gk + \geta] \eolnn
&\qquad\quad - \frac 5{9\sqrt 6} \scr H [\gpi + \gk] \eolnn
&\qquad\quad + \frac 1{3\sqrt 6}[(D - 3F)\ggpi
+ 3(D - F)\ggk + 2D\ggeta], \eol
\gam_{n\Sig^-}^{4 + i5} &= -\frac 1{12}(D - F)
[14\gpi + 13\gk + 3\geta] \eolnn
&\qquad\quad + \frac 5{81} \scr H [2\gpi + \gk] \eolnn
&\qquad\quad + \frac 19[2(D + 5F)\ggpi
+ (D + 5F)\ggk - (D - 3F)\ggeta], \eol
\gam_{\Sig^0\Xi^-}^{4 + i5} &= -\frac 1{12\sqrt 2} (D + F)
[5\gpi + 19\gk + 6\geta] \eolnn
&\qquad\quad - \frac 5{81\sqrt 2} \scr H [2\gpi + 7\gk + 3\geta] \eolnn
&\qquad\quad + \frac 1{9\sqrt 2}[2(2D + F)\ggpi
+ (15D + 13F)\ggk \eolnn
&\qquad\qquad\qquad\qquad + 3(D + F)\ggeta], \eeol
\eeq
where
\eqa
G_1(m) &\equiv \left( m^2-2\Del^2\right) \ln{m^2\over\mu^2}
+ 4 m \Del \left(1 - \frac{\Del^2}{m^2}\right) F(\Del / m), \eol
G_2(m) &\equiv \left(m^2-{2\over 3}\Del^2\right) \ln{m^2\over\mu^2}
- \frac{4m^3}{3\Del}
\left[\left(1 - \frac{\Del^2}{m^2}\right)^2 F(\Del / m)
- \frac\pi 2 \right] - \frac 43 m^2. \eeol
\eeq
Here we have defined
\eq
F(x) \equiv \cases{\displaystyle
\frac 1{\sqrt{1 - x^2}} \tan^{-1}\frac{\sqrt{1 - x^2}}x
& for $x \le 1$, \cr\displaystyle
\frac 1{2\sqrt{x^2 - 1}} \ln\frac{x + \sqrt{x^2 - 1}}{x - \sqrt{x^2 - 1}}
& for $x > 1$. \cr}
\eeq

To obtain these results, we have dropped terms analytic in the quark masses
which can be absorbed into counterterms in the chiral lagrangian.
This amounts to a choice of subtraction procedure.
The limiting values of the decuplet corrections can be obtained using
\eqa
G_1(m) &= \cases{\displaystyle
m^2\ln\frac{m^2}{\mu^2} & for $\Del \ll m$, \cr\displaystyle
\left(m^2-{2\over 3}\Del^2\right) \ln\frac{4\Del^2}{\mu^2} -
m^2\phantom{\frac 53}
& for $\Del \gg m$, \cr} \eol
G_2(m) &= \cases{\displaystyle
m^2 \ln\frac{m^2}{\mu^2} & for $\Del \ll m$, \cr\displaystyle
\left(m^2-{2\over3}\Del^2\right) \ln\frac{4\Del^2}{\mu^2} - \frac 53 m^2
& for $\Del \gg m$. \cr} \eeol
\eeq

From the expressions above it is easy to check that the decuplet contributions
decouple in the limit $\Del \gg m$.
In this limit the decuplet corrections are either analytic in the quark masses,
and can be absorbed into terms in the effective lagrangian which contain
only octet baryon fields, or are $SU(3)$ symmetric $O(\Del^2 \ln \Del^2)$
terms which can be absorbed into a redefinition of $D$ and $F$.
We can also consider the limit $\Del \ll m_K, m_\eta$ advocated in
ref.\ \JMdec.
We find that $G_1(m)$ and $G_2(m)$ are very poorly approximated by setting
$\Del = 0$: for example, they have the wrong sign.

The decuplet corrections are large, $\sim 100\%$, like the octet corrections.
However, unlike the octet corrections, wavefunction renormalization is
not the largest part of the decuplet corrections.
This means that these corrections cannot be removed by an
$SU(3)$-conserving field redefinition, as contemplated above.

We now compare our results to semileptonic decay data (see the appendix).
The constant $\scr C$ can be determined from non-leptonic weak decays
of decuplet states to be $\scr C \simeq 1.5$ \JMdec, so there is one new
undetermined parameter $\scr H$ compared to the octet case.
We assign a $20\%$ theoretical uncertainty due to $O(m_s)$ corrections
to the amplitudes and show the resulting $67\%$, $90\%$, and $95\%$
confidence level region in the $D$, $F$ plane in fig.\ 1.
(The region is obtained by projecting the allowed region in $D$, $F$,
and $\scr H$ space onto the $D$, $F$ plane.)
Because of the large allowed region, a best fit is probably
meaningless.

Despite the fact that the allowed region is quite large, one can
draw some non-trivial conclusions.
The first is that the lowest-order values of $D$ and $F$ are allowed.
For these values of $D$ and $F$, the corrections are less than $10\%$
if $\scr H \simeq 0.5$.
(A similar cancellation was found in ref.\ \JMdec, but for a different
range of $\scr H$ values.)
This suggests that including decuplet baryon states may improve
the convergence of baryon chiral perturbation theory.

We emphasize that the fact that the decuplet contributions can cancel
the octet contributions is highly non-trivial:
it involves a large cancellation between octet wavefunction
renormalization and decuplet ``vertex'' corrections.
(The decuplet contribution to wavefunction renormalization is positive
and therefore cannot cancel the octet wavefunction renormalization
contribution.)
Also, it is striking that the cancellation occurs both for nonzero as
well as zero octet--decuplet mass splitting, since the corrections are
very different in these two cases.

Finally, we can ask whether the corrections favor the $SU(6)$ prediction
\eq
F = \sfrac 23 F, \qquad
\scr C = -2 D, \qquad
\scr H = -3 D.
\eeq
This relation is excluded by the above analysis at the $95\%$ confidence
level.
(If we set $\Del = 0$, the corrections favor the $SU(6)$ values, as
found in ref.\ \JMdec.)
It is not clear how meaningful this is, since higher order corrections
may be important.

\section{Conclusions}

We have critically examined the chiral perturbation theory predictions
for the axial vector form factors, both with and without the inclusion
of decuplet intermediate states.
We confirmed the result of ref.\ \JMax\ that the corrections are large
if decuplet states are not included.
We argued that these large corrections are a symptom that chiral
perturbation theory including only octet baryon states is breaking down,
despite the fact that the largest correction takes the
form of an $SU(3)$-singlet wavefunction renormalization.
We then examined the contributions of decuplet intermediate states.
We found that taking into account the effects of the decuplet--octet
mass difference substantially changes the pattern of corrections
obtained in ref.\ \JMdec, which neglects these effects.
We found that the decuplet corrections tend to cancel the octet
corrections in a non-trivial way (as also found by ref.\ \JMdec), and
that the corrections strongly disfavor the $SU(6)$ values of the axial
couplings (contrary to the conclusions of ref.\ \JMdec).
We also argued that $D$ and $F$ cannot be reliably extracted from a
calculation which includes only the non-analytic corrections.

While completing the present paper, we received
ref.\ \ref\them{V. Bernard, N. Kaiser and Ulf-G. Mei\ss ner, BUTP-93/05,
CRN-93-06, (hep-ph/9303311), unpublished.},
which also makes the point (in a different context) that setting
$\Del = 0$ in the decuplet integrals is not a good approximation.

\section{Acknowledgements}

We would like to thank E. Jenkins, M. Luke, A. V. Manohar, M. Savage, and
M. Suzuki for discussions.
This work was supported by the Director, Office of Energy Research, Office
of High Energy and Nuclear Physics, Division of High Energy Physics of the
U.S.\ Department of Energy under Contract DE-AC03-76SF00098.

\appendix{A}{Fit to Semileptonic Decays}

In this appendix, we consider the determination of $D$ and $F$ from
$\Delta S = 1$ semileptonic decays of hyperons.
These decays are governed by the form factors
\eqa
\bra{B_a} J^{V}_{\mu c}(0) \ket{B_b} &= \mybar u(p_a) \biggl[
f_1^{abc}(q^2) \gamma_\mu
+ \frac{if_2^{abc}(q^2)}{M_a + M_b} \sigma_{\mu\nu} q^\nu
+ \frac{if_3^{abc}(q^2)}{M_a + M_b} q_\mu \biggr] u(p_b), \eol
\bra{B_a} J^{A}_{\mu c}(0) \ket{B_b} &= \mybar u(p_a) \biggl[
g_1^{abc}(q^2) \gamma_\mu \gamma_5
+ \frac{ig_2^{abc}(q^2)}{M_a + M_b} \sigma_{\mu\nu} \gamma_5 q^\nu
+ \frac{ig_3^{abc}(q^2)}{M_a + M_b} \gamma_5 q_\mu \biggr] u(p_b),
\ \ \eeol
\eeq
where $q \equiv p_a - p_b$.
The contributions of the form factors $f_3$ and $g_3$ are suppressed by
the electron mass, and can be safely neglected.
Near the $SU(3)$ limit $m_u = m_d = m_s$, the baryons are nearly
degenerate, and at the order we are working the decays are determined by
the form factors at zero momentum transfer.
The contributions of $f_2$ and $g_2$ are suppressed by $O(m_s)$ because
of the explicit power of $q$ multiplying these terms.
(In fact, time-reversal invariance can be used to show that $g_2(0) = 0$
in the $SU(3)$ limit, so that the contributions of $g_2$ are even
smaller.)
In the $SU(3)$ limit, the form factors $f_1(0)$ are given by
Clebsch--Gordan coefficients and the $g_1(0)$'s are simple linear
combinations of $D$ and $F$ (see eq.\ (13)).
The corrections to $f_1$ are $O(m_s)$ and can be computed in chiral
perturbation theory \ourvec.
The leading corrections to $g_1$ are $O(m_s \ln m_s)$ and formally
give the largest corrections to the semileptonic decay rates.
These are therefore the only corrections to the form factors which
we will keep.

To perform our fit, we use both decay rate and asymmetry data taken from
the most recent Particle Data Group (PDG) compilation
\ref\PDG{Particle Data Group, \PRD{45}{S1}{1992}.}.
For the asymmetry data, we directly use the average values for
$g_A / g_V$ quoted by the PDG.
To convert the decay rates into values for $g_1$, we keep the full
kinematic dependence on the baryon masses, since these effects turn  
out to be numerically important.
The data we use is displayed in table 1.

\def \phm {\phantom{-}}
\vbox{ \vskip 20pt \centerline{
\vbox{ \offinterlineskip
\halign {\vrule#& \hfil#\hfil& \vrule#& \hfil#\hfil& \vrule#&
\hfil#\hfil& \vrule# \cr
\noalign{\hrule}
height2pt& \omit&& \omit&& \omit& \cr
&\qquad\qquad\qquad&& \qquad lifetime\qquad&& \qquad asymmetry\qquad&\cr
height2pt&\omit&&\omit&&\omit&\cr \noalign{\hrule}
height2pt&\omit&&\omit&&\omit&\cr
&$n^{\phm}\to p^{\phm}$&&
$\phm1.323 \pm 0.003$ && $\phm1.257\pm0.003$ &\cr
height2pt&\omit&&\omit&&\omit&\cr \noalign{\hrule}
height2pt&\omit&&\omit&&\omit&\cr
&$\Sig^-\to\Lam^{\phm}$&& $\phm0.609\pm0.029$ && $\phm0.62\pm0.44$ &\cr
height2pt&\omit&&\omit&&\omit&\cr \noalign{\hrule}
height2pt&\omit&&\omit&&\omit&\cr
&$\Lam^{\phm}\to p^{\phm}$&& $-0.972\pm0.018$ && $-0.879\pm0.021$ &\cr
height2pt&\omit&&\omit&&\omit&\cr \noalign{\hrule}
height2pt&\omit&&\omit&&\omit&\cr
&$\Sig^-\to n^{\phm}$&& $\phm0.442\pm0.021$ && $\phm0.340\pm0.017$ &\cr
height2pt&\omit&&\omit&&\omit&\cr \noalign{\hrule}
height2pt&\omit&&\omit&&\omit&\cr
&$\Xi^- \to\Sig^0$&& $\phm0.96\phantom{0}\pm0.19\phantom{0}$ && ------ &\cr
height2pt&\omit&&\omit&&\omit&\cr \noalign{\hrule}
height2pt&\omit&&\omit&&\omit&\cr
&$\Xi^- \to \Lam^{\phm}$&& $\phm0.473\pm0.026$ && $\phm0.306\pm0.061$ &\cr
height2pt&\omit&&\omit&&\omit&\cr \noalign{\hrule} }} } \vskip 10pt }
\centerline{Table 1: Values for $g_1(0)$ extracted from 1992 PDG}

The decay rate and asymmetry determinations of $g_1$ are inconsistent
if we assume only the errors quoted by the PDG.
This is either a symptom of systematic errors in the experiments or an
indication that higher-order corrections are important.
We expect that higher order terms in the chiral expansion will give
rise to $\sim 20\%$ corrections, and so we added this amount in quadrature
to all the quoted errors to take into account the theoretical uncertainty.
When we do this, all the errors on all determinations have a
sizable overlap.

With this procedure, reasonable fits are obtained.
For example, if we fit this data to $D$ and $F$ using the lowest-order
prediction in eq.\ \loword, we obtain the best fit
\eq
\label\lowfit
D = 0.85\pm 0.06, \qquad
F = 0.52\pm 0.04,
\eeq
with $\chi^2 = 6.1$ for 9 degrees of freedom.

\listrefs

\vfill
\eject
\centerline{\bf Figure Captions}
\vskip .4in
\noindent
Fig.\ 1.
Contours of $68\%$, $90\%$, and $95\%$ confidence-level regions in
the $D$--$F$ plane when decuplet corrections are included, obtained as
discussed in the text.
The black dot shows the lowest-order values of $D$ and $F$, as discussed
in the appendix.

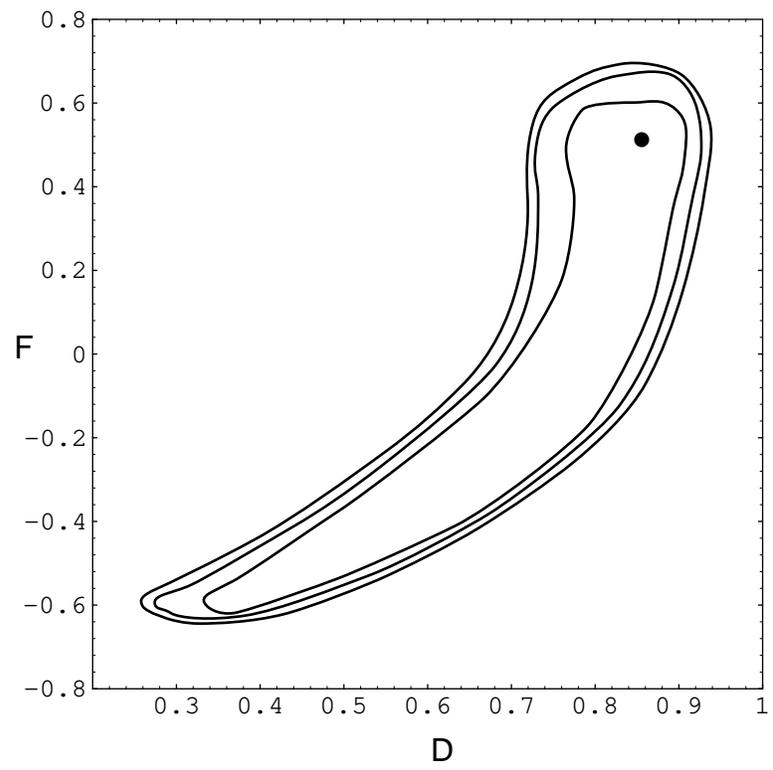
\bye